\documentclass[pre,showpacs,twocolumn]{revtex4}

\usepackage{amsmath}
\usepackage{graphicx}
\usepackage{dcolumn}
\def \dif {\mathrm{d}}
\def \erf {\mathrm{erf}}
\begin{document}

\title{A Monte Carlo simulation of ion transport at finite temperatures}

\author{Zoran Ristivojevic}
\affiliation{Laboratoire de Physique Th\'{e}orique-CNRS, Ecole Normale Sup\'{e}rieure, 24 rue Lhomond, 75005 Paris, France}
\affiliation{Institute of Physics, University of Belgrade, PO Box 68, 11080 Zemun, Belgrade, Serbia}

\author{Zoran Lj.~Petrovi\'{c}}%
\affiliation{Institute of Physics, University of Belgrade, PO Box 68, 11080 Zemun, Belgrade, Serbia}

\date{\today}

\begin{abstract}
We have developed a Monte Carlo simulation for ion transport in hot background gases, which is an alternative way of solving the corresponding Boltzmann equation that determines the distribution function of ions. We consider the limit of low ion densities when the distribution function of the background gas remains unchanged due to collision with ions. A special  attention has been paid to properly treat the thermal motion of the host gas particles and their influence on ions, which is very important at low electric fields, when the mean ion energy is comparable to the thermal energy of the host gas. We found the conditional probability distribution of gas velocities that correspond to an ion of specific velocity which collides with a gas particle. Also, we have derived exact analytical formulas for piecewise calculation of the collision frequency integrals.  We address the cases when the background gas is monocomponent and when it is a mixture of different gases. The developed techniques described here are required for Monte Carlo simulations of ion transport and for hybrid models of non-equilibrium plasmas. The range of energies where it is necessary to apply the technique has been defined. The results we obtained are in excellent agreement with the existing ones obtained by complementary methods. Having verified our algorithm, we were able to produce calculations for Ar$^+$ ions in Ar and propose them as a new benchmark for thermal effects. The developed method is widely applicable for solving the Boltzmann equation that appears in many different contexts in physics.
\end{abstract}


\maketitle

\section{\label{sec:level1}Introduction}

Non-equilibrium plasmas are used in a broad range of applications
that include plasma etching
\cite{MP06,LL05}, biomedical applications
\cite{SS05,ZPGGP04}, nanotechnologies
\cite{KWM04,PKL07} and microdischarges \cite{SS99}. Development of devices based on non-equilibrium plasmas is quite expensive and cannot be based on empirical methods, therefore predictive accurate models had to be developed \cite{KCGHB96,MP06}.

Ion transport coefficients are used as input parameters in both fluid and hybrid models of plasmas
\cite{MP06,RWP05,BP91,SGJ90,VSHK93,D01}
relevant for the applications mentioned above. Indirectly, transport
coefficients can be used to verify the completeness and the absolute magnitude of the cross sections in the sets \cite{RWP05,MP06} that are to be used in Monte Carlo (MC) and particle in cell models of
plasmas \cite{K83,PBJD95}. There is a critical shortage of data for most ions in most gases that are interesting for applications \cite{PRJSMMU07}. In addition, there is a pressing need to have both accurate and efficient algorithms for simulation of ion motion. For example, even in hybrid models, that were developed to take into account the non-local transport of electrons, more and more frequently ions and even fast neutrals are treated by an MC technique \cite{D01,GBG03,GBG04,MHMDP03}. The procedure developed in the present paper is quite general for all externally driven particles in background gases that are at non-zero temperatures. We primarily focus on ions because they have energies close to the thermal energy over a wider range of reduced electric fields, although the same effects of the motion of background particles exist for electrons and and neutral particles.

The standard theory of ion transport has been well established and has become a part of textbooks \cite{MM88}. More recently some of the important developments have been achieved, which include application of transport theory to gas-filled ion traps \cite{viehland+10}, radio-frequent transport of ions \cite{white01,robson+94}, general discussions of transport in crossed electric and magnetic fields \cite{dujko+11} and transport data and cross sections \cite{benhenni+09}.

\subsection{Boltzmann equation and the distribution function}

In the following we will consider ions that are driven by an external electric field through the neutral background gas. Their transport coefficients can be easily calculated once the distribution function is known. The latter is determined by the Boltzmann equation (BE)\cite{Wannier} for ions--neutral gas collisions, which reads
\begin{align}\label{BE}
&\frac{\partial f(\mathbf{x},\mathbf{v},t)}{\partial t}+\mathbf{v}\nabla_x f(\mathbf{x},\mathbf{v},t)+\frac{\mathbf{F}_e}{m}\nabla_v f(\mathbf{x},\mathbf{v},t)\notag\\
&=\int\sigma^d |\mathbf{v}-\mathbf{u}|
\left[f(\mathbf{x},\mathbf{v}',t)F(\mathbf{u}') -f(\mathbf{x},\mathbf{v},t)F(\mathbf{u})\right]\dif\Omega\dif\mathbf{u}.
\end{align}
Here $f(\mathbf{x},\mathbf{v},t)$ and $F(\mathbf{u})$ are the distribution functions of ions and of the neutral gas respectively, $m$ is the mass of ions, $\dif\Omega=\sin\theta\dif\theta\dif\psi$ is the solid angle and $\sigma^d(|\mathbf{v}-\mathbf{u}|,\theta)$ is the total differential cross section for ion-neutral gas collisions. The angle $\theta$ denotes the angle of rotation of the relative velocity of collision particles before and after the collision. The postcollision velocities $\mathbf{v}'$ and $\mathbf{u'}$ are determined by the momentum and energy conservation laws from the precollision velocities $\mathbf{v}$ and $\mathbf{u}$. The external force is denoted by $\mathbf{F}_e$. In (\ref{BE}) we assumed that the neutral gas is homogeneous and that its distribution function does not significantly change due to collisions with ions.

The collision integral of the BE [right hand side of (\ref{BE})] is a functional that depends on the distribution functions of both, ions and the background gas and also depends on the total cross section for their collisions. Since realistic cross sections are almost always complicated functions of the relative velocity ion-neutral gas, direct solution of the BE is often quite complicated, apart from some model cross sections. Instead of solving the BE an alternative route in determining the ion distribution function and  their transport parameters (e.g.~the mean energy, the drift velocity, the diffusion coefficients) is by an MC simulation where one simulates real physical processes and follows the temporal evolution of ions. Since the initial conditions for ions are arbitrary, the system should evolve for some time until the stationary regime is achieved, when the distribution function of ions becomes time independent. After the system has relaxed, it is quite simple task to directly sample all ion properties in a simulation. Due to stochastic nature of the MC method, an average over many uncorrelated measurements is necessary.

The first attempt to apply an MC simulation for ions in
gases was made by Wannier \cite{W53}. Here we will briefly describe the technique. In the simulation we follow the evolution of a certain number of ions that collide with the background gas. For low ion densities, the interaction among ions is usually neglected and only the interaction with the host gas is taken into account. Different atomic processes between ions which collides with host particles are taken into account through the set of cross sections. The total cross section defines the collision frequency $\nu(v)$, which is a function of the ion velocity $v$. The time between successive collisions of an ion with host particles is stochastically determined knowing the the collision frequency. Since for low densities of host gases, the ion mean free path is much larger than the thermal de Broglie wavelength (we consider not too low temperatures), the ion propagates between collisions according to the laws of classical mechanics. Consequently, quantum mechanics is contained only in the cross sections. The act of collision with a host particle is treated using the conservation laws and also the information from the differential cross sections about possible anisotrophies of the direction of an ion after the collision. More details about the MC technique one could find, for example, in a review paper \cite{N00}.

When calculating properties of ions that move in a neutral host gas using MC simulation \cite{PS98} or momentum transfer theory \cite{JVP02}, it is very often assumed that the particles of the neutral gas are at rest. It is a reasonable approximation when the mean ion energy significantly exceeds the
thermal gas energy. Then the collision frequency takes a simple form (\ref{ni_coldgas}). However, this approximation is
not justified for low electric fields, when the mean ion energy is comparable to the thermal energy. In such case the motion of the host gas particles should be taken into account and the collision frequency has to be calculated more accurately that takes into account the thermal motion of gas particles. At this point we want to emphasize that in MC simulations we only follow the evolution of ions in time, since it is assumed that their motion does not significantly change the distribution function of background gas particles, i.e. the gas heating effects are neglected. All properties of the host gas particles are determined by their distribution function.
If however, there is a significant gas heating due to current flow, which is relevant for higher current discharges, the background gas may be at a considerably higher temperature. This would only extend the need to apply our technique but in that case the calculation of the temperature would have to be done in a self consistent manner according to the experimental conditions.

One of the first papers that addressed the question of thermal effects of the background gas was the one of Lin and Bardsley \cite{LB77}. They used an approximate treatment and sampled the velocities of the host gas directly from the gas distribution function when an ion collided with a host particle. This procedure is not entirely correct, since during the process of MC simulation, at the moment when an ion of velocity $\mathbf{v}$ collides with a host gas particle, the probability that it collides with the gas particle of velocity $\mathbf{u}$ is not only a function of the gas distribution function, but also depends on the cross section and the relative velocity [see (\ref{gustinaverovatnocesudara})]. Using the gas distribution function for sampling the gas velocities is an approximation that leads to systematic errors in results, as we demonstrate explicitly on an example, see Sec. \ref{ccs} and Fig.~\ref{slikatrcoef}. The most dramatic consequences of wrong sampling of the background gas velocities is the absence of equilibration at vanishing electric field as well as wrong ratio between the diffusion coefficient and mobility.

The rest of the paper is organized in the following way. In section II we derive and analyze an explicit expression for the collision frequency. We derive an asymptotic formula for the collision frequency which is valid for large ion velocities with respect to the background gas thermal velocity. In
section III we emphasize a proper way of sampling velocities of the background gas, define the corresponding probability functions and propose the rejection methods for sampling background gas velocities from that distribution. The rejection methods depend on the specific form of cross section and we considered the most common two cases. In section IV we describe the procedure for thermal gas effects in the case of mixtures of gases. In section V we describe some details of the MC simulation. Numerical results, comparison with existing data and benchmark results are given in section VI which is followed by conclusions.

\section{Collision frequency}

The collision frequency for an ion of velocity $\mathbf{v}$ in a neutral background gas is \cite{Wannier}
\begin{align}
\label{nidef} \nu(v)=n\int\sigma
\left(|\mathbf{v}-\mathbf{u}|\right) {|\mathbf{v}-\mathbf{u}|}
F(\mathbf{u})\dif\mathbf{u},
\end{align}
where $n$ is the gas density and $\sigma
\left(|\mathbf{v}-\mathbf{u}|\right)$ is the total cross section for ion--neutral gas scattering processes as a function of relative velocity, obtained by integrating the total differential cross section introduced in (\ref{BE}) over the solid angle. Having in mind that, the collision frequency should be calculated for many different ion velocities and that the form of cross sections is in general a complicated expression, one needs to be able to calculate (\ref{nidef}) very efficiently. In addition, one should be aware that ions have thermal energies for a very wide range of reduced fields covering most practical discharges.

In principle, equation (\ref{nidef}) is a simplified in the sense that at this point we deal with the total cross section. However, the point that (\ref{nidef}) contains the total cross section rather than the differential present in (\ref{BE}) is due to the fact that the angular dependence in $\sigma^d(|\mathbf{v}-\mathbf{u}|,\theta)$ could be integrated out from quantities that do not depend on the postcollision velocities. While the BE (\ref{BE}) does depend on the latter, and we have to deal with the differential cross section, the collision frequency does not and we could deal with the total cross section. The angular dependence determined by $\sigma^d$ is taken into account later for the kinematics, when the angle of scattering is determined. In some systems when the force is not central a more complex form of dependence on the vector of the relative speed may be required. However, the complex form of interaction is often not found in the species used for processing plasmas. In this paper we use the model of anisotropy proposed by Phelps \cite{P94} for argon ions which is not as detailed as using complete differential cross section but is sufficiently simple to be applied in plasma modeling and yet it reproduces all the important features of ion transport with high accuracy. However, our approach is not limited to that simplified approach and one could use arbitrary differential cross sections.

Different approaches for the calculation of the collision frequency can be found in the literature. In \cite{LD04} calculation of the integral in (\ref{nidef}) is performed by numerical integration using an MC integration technique. While being simple, this technique is not a very efficient algorithm, and its accuracy could be improved at the expense of the efficiency. Another approach is proposed in \cite{RS03} where authors treat the problem in a mean field like approach. They define the mean relative velocity, which becomes the argument of the cross section. Taking the cross section outside the integral, the remaining integral in (\ref{nidef}) is easily doable. In this paper we have set out to find an accurate and efficient algorithm which may be used for MC simulations of a ions (as well as electrons). In doing so we should remember that most cross sections are defined numerically at a limited number of points and that linear interpolation between those points is often used. Using that we have developed an MC simulation which is used to  propose benchmark calculations for ion motion in their parent gas at non-zero temperatures.

Explicit calculation of the collision frequency (\ref{nidef}) is only possible for specific cross
sections and gas velocity distribution functions. In the following we will assume that the background gas is in equilibrium at temperature $T$, so that is
described by a Maxwell velocity distribution function
(MVDF)
\begin{align}
F(\mathbf{u})=\frac{1}{(\pi w^2)^{3/2}}\exp\left({-\frac{\mathbf{u}^2}{w^2}}\right),
\end{align}
where the most probable gas thermal velocity is
$w=\left(2kT/M\right)^{1/2}$, $M$ is the mass of the gas particle, $k$ is the Boltzmann constant and $T$ the temperature. After angular integration, expression (\ref{nidef}) transforms into
\begin{align}
\label{def_kolizionav} \nu(v)=&\frac{n}{\sqrt\pi
wv}\int_{0}^{+\infty}\dif x\sigma(x)x^2\notag\\
&\times\Bigg\{\exp\left[-\frac{(x-v)^2}{w^2}\right]
-\exp\left[-\frac{(x+v)^2}{w^2}\right]\Bigg\}.
\end{align}
A similar expression may be derived for the ion velocity identically equal to zero, but this case is not encountered in realistic simulations and will not be considered explicitly here. One can alternatively determine $\nu(0)$ as the limiting case $v\rightarrow 0$ of (\ref{def_kolizionav}).

It is possible to convert the integral of the type
(\ref{def_kolizionav}), to give the following expression for the
collision frequency when ion velocities are much gre\-a\-ter than
the thermal gas velocity (see appendix for details):
\begin{align}
\label{ni_vveliko}
\nu(v)=n\sigma(v)v\left(1+\frac{w^2}{v^2}\frac{[v^2\sigma(v)]''}
{4\sigma(v)}+\ldots\right)
\end{align}
We see that the first order correction is proportional to the square of the ratio $w/v$, which determines the criterion for importance of the thermal background gas motion. The collision frequency for cold gas ($w=0$) is a
special case of (\ref{ni_vveliko}), when we get
\begin{align}
\label{ni_coldgas} \nu(v)=n\sigma(v)v.
\end{align}
This formula is very widely used in MC simulations, for
example in \cite{B91,VS95,PS98} even though it is not applicable at low ion energies and at nonzero temperatures. Here we mention that the range of applicability of the previous expressions for the collision frequency is the same as for the standard transport theory with well isolated single collisions. At very high pressures multiple collisions may become important and treatment of such processes requires additional assumptions.

For further evaluation of (\ref{def_kolizionav}) we need a
specific form of the cross section. Realistic cross sections are always
tabulated as a function of incident ion energies (or velocities) or
relative energies (velocities).  In principle it could be argued
that the cross sections are often tabulated as constant values in
narrow energy bins so the solution for a constant cross section may
be sufficient.  However in the low energy limit the cross sections
often increase rapidly and application of a constant cross section
may require application of a large number of energy bins. At the
same time usually the data for the low energy limit of mobilities is
available and thus one needs accurate and efficient schemes for
calculation in this energy range. Thus we assume quite generaly that the cross section, given as a function of the relative ion-neutral velocity $x$, may be expressed as
\begin{align}
\label{presek} \sigma(x) = \left\{ \begin{array}{cc}
a_1^2\left(\frac{x}{v}\right)^2+a_1^1\left(\frac{x}{v}\right)+a_1^0, & 0\leq x\leq x_1\\
a_2^2\left(\frac{x}{v}\right)^2+a_2^1\left(\frac{x}{v}\right)+a_2^0, & x_1\leq x\leq x_2\\
\vdots & \vdots\\
a_i^2\left(\frac{x}{v}\right)^2+a_i^1\left(\frac{x}{v}\right)+a_i^0, & x_{i-1}\leq x\leq x_{i}\\
\vdots & \vdots
\end{array} \right.
\end{align}

This expansion is merely a polynomial expansion used for tabulation
of the data in the code extending the usually implemented procedure
to use constant cross sections in each of the energy bins. This
expansion allows three terms which all may provide analytic
solutions later to facilitate the speed of implementation.  The
expansion is intended to be used in (2) and here $x$ is the relative
velocity while $v$ is the velocity of the ion.

Taking into account (\ref{presek}) we can define integrals which
originate in (\ref{def_kolizionav}) as:
\begin{align}
\label{i0def}
I_{j}^{\pm}(v,w,x_0,x_{1})=\int_{x_0}^{x_{1}}\dif x\frac{x^{2+j}}{\sqrt\pi
wv^{2+j}}\exp\left[-\frac{(x\pm v)^2}{w^2}\right],
\end{align}
with $j=0,1,2$. These integrals can be calculated analytically, and
in terms of the variables $a_i^\pm=(x_i\pm v)/w$ they read
\begin{widetext}
\begin{align}
\label{i0}
I_0^{\pm}(v,w,x_0,x_{1})=\frac{1}{4}\left(2+\frac{w^2}{v^2}\right)
\left[\erf(a_1^\pm)- \erf(a_0^\pm)\right]
+\frac{1}{2\sqrt\pi}\frac{w^2}{v^2}
\left[a_0^\mp\exp(-{a_0^\pm}^2)-a_1^\mp\exp(-{a_1^\pm}^2)\right],
\end{align}
\begin{align}
\label{i1}
I_1^{\pm}(v,w,x_0,x_{1})=&\mp\frac{1}{4}\left(2+3\frac{w^2}{v^2}
\right) \left[\erf(a_1^{\pm})-\erf(a_0^{\pm})\right]
+\frac{1}{2\sqrt\pi}\exp(-{a_0^{\pm}}^2)
\left[3\frac{w}{v}\mp3\frac{w^2}{v^2}a_0^{\pm}
+\frac{w^3}{v^3}\left(1+{a_0^{\pm}}^2\right) \right]\notag\\
&-\frac{1}{2\sqrt\pi}\exp[-{a_1^{\pm}}^2]
\left[3\frac{w}{v}\mp3\frac{w^2}{v^2}a_1^{\pm}
+\frac{w^3}{v^3}\left(1+{a_1^{\pm}}^2\right)\right],
\end{align}
\begin{align}
\label{i2}
I_2^{\pm}(v,w,x_0,x_{1})=&\frac{1}{8}\left(4+12\frac{w^2}{v^2}+
3\frac{w^4}{v^4}\right)
\left[\erf(a_1^{\pm})-\erf(a_0^{\pm})\right]\notag\\
&+\frac{1}{4\sqrt\pi}\exp(-{a_0^{\pm}}^2)
\bigg[\mp8\frac{w}{v}+12\frac{w^2}{v^2}a_0^{\pm}
\mp8\frac{w^3}{v^3}\left(1+{a_0^{\pm}}^2\right)
+\frac{w^4}{v^4}a_0^{\pm}\left(3+2{a_0^{\pm}}^2\right)\bigg]\notag\\
&{-\frac{1}{4\sqrt\pi}\exp(-{a_1^{\pm}}^2)
\bigg[\mp8\frac{w}{v}+12\frac{w^2}{v^2}a_1^{\pm}} \mp
8\frac{w^3}{v^3}\left(1+{a_1^{\pm}}^2\right)
+\frac{w^4}{v^4}a_1^{\pm}\left(3+2{a_1^{\pm}}^2\right)\bigg].
\end{align}
Now, the collision frequency (\ref{def_kolizionav}) can be easily
calculated as
\begin{align}
\label{collfreq-eq10}
\nu(v)=\sum_{j=0}^{2}\sum_{i=0}^{\infty}nva_i^j
\left[I_j^-(v,w,x_i,x_{i+1})-I_j^+(v,w,x_i,x_{i+1})\right].
\end{align}
\end{widetext}

We notice that this sum is made over all values of the index $i$,
i.e.~for all relative velocities $x$, but in practical applications
the situation is different. Integrals (\ref{i0def}) are homogenous
functions of their arguments of the order zero, and we will now
consider some of their general properties. Under the conditions $v\geq 6w,\quad x_0=0,\quad  x_1=\infty,$ the integrals satisfy
\begin{align}
\label{Iplusv6w}
&I_0^+,\ I_1^+,\ I_2^+<10^{-20},\\
\label{Iminusv6w}
 &I_0^-,\ I_1^-,\ I_2^-\approx 1.
\end{align}
Also, under the conditions $v\geq 6w,\quad x_0\geq v+6w,\quad  x_1\geq x_0, $
the integrals satisfy
\begin{align}
I_0^-,\ I_1^-,\ I_2^-<2\cdot10^{-16}.
\end{align}
In other words, the integrals decrease very
rapidly as $v$ increases.

For a double precision when the machine accuracy is $\sim 10^{-16}$ and a given ion velocity $v\geq
6w$, it is enough to integrate over the relative velocities up to $v+6w$. Moreover, because of (\ref{Iplusv6w}) and (\ref{Iminusv6w}), the
expansion (\ref{ni_vveliko}) converges very rapidly.
For ion velocities which are less than $6w$ similar consideration can be made. One can conclude that the upper limit for integration in (\ref{i0def}) should be at most $12w$.

\subsection{Constant cross section}

For a constant cross section $\sigma_0$, from
(\ref{def_kolizionav}), (\ref{presek}) and (\ref{i0}) we obtain the
collision frequency for  hard sphere scattering (which was
previously found and used in \cite{NW99}):
\begin{align}
\label{nivhardfinal}
\nu(v)=&n\sigma_0w\Biggl\{\left(\frac{v}{w}+\frac{w}{2v}\right) \erf\left(\frac{v}{w}\right)\notag\\
&+\frac{1}{\sqrt\pi}\exp\left[-\left(\frac{v}{w}\right)^2\right]\Biggr\}.
\end{align}
Either using (\ref{ni_vveliko}) or expanding the previous equation for small ratio $w/v$, the collision frequency (\ref{nivhardfinal}) may be approximated by
\begin{align}
\label{nivhardfinalapprox}
\nu(v)\simeq n\sigma_0v\left(1+\frac{w^2}{2v^2}\right).
\end{align}
One can notice that, according to (\ref{ni_vveliko}), in the case of a constant cross section, only the the first order power law correction exists.

\subsection{Realistic cross sections}

\begin{figure}[t]
\centering
\includegraphics[width=\linewidth]{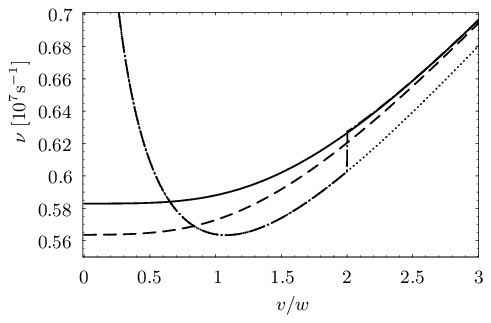}
\caption{The collision frequency for Ar$^+$ ions in their parent gas for
a cross sections from \cite{P94} for the parameter $m=2$. The gas
density is $n=10^{22}\;\mathrm{m}^{-3}$ and the temperature
$T=77\;\mathrm{K}$. The solid line is calculated using the exact formula
(\ref{def_kolizionav}) or (\ref{collfreq-eq10}), the dashed line using
the approximate formula from \cite{RS03}, the dash--dotted by using our expression (\ref{ni_vveliko}) with inclusion of the correction terms after $2w$, while the dotted curve is determined from the cold gas approximation (\ref{ni_coldgas}). The dashed and the dash-dotted curve overlap for $v<2w$. The cold gas approximation significantly overestimates the collision frequency at small velocities and leads to wrong results.}\label{slikaargon}
\end{figure}

Calculations of the collision frequency are also done for a realistic cross section \cite{P94} for Ar$^+$ ions in their parent gas. In Fig.~\ref{slikaargon} we show the collision frequency calculated using different methods. The exact calculation by using the functional form of the cross section from \cite{P94} in the definition (\ref{nidef}) and the calculation based on equation (\ref{collfreq-eq10}) are essentially identical on the scale of the Figure. We can see that the approximate formula from \cite{RS03} (the dashed line) should be improved further (toward the solid line). The formula (\ref{ni_coldgas}) for cold gas approximation (the dash-dotted line) is obviously inaccurate. However it could be cured by the the analytic
correction in (\ref{ni_vveliko}) (for illustration, the correction is showed $v>2w$). Relative
difference between the exact formula (\ref{def_kolizionav}) and formula (\ref{ni_vveliko}) is less than $1.3\cdot10^{-3}$ for
$v>2w$, and less than $7.2\cdot10^{-6}$ for $v>6w$. Having in mind that the relative difference between the exact formula and the cold gas approximation (\ref{ni_coldgas}) even at $v=18w$ is greater than $10^{-3}$ and increases with decreasing $v$, we may conclude that the corrected form of calculation is necessary. For large  $v/w$
limit and in order to save the computational time the analytic form (\ref{ni_vveliko}) of the correction term is very useful and sufficiently accurate for all practical purposes.

The main point of this paper is to provide a technique to treat thermal collisions and then to provide benchmark results to test other codes.  Thus we do not enter detailed discussions on the ion transport in argon and we do not discuss relative merits of different sources or experimental data or the analysis that ensued as we could add very little to that which is provided in \cite{P94,MM88} and other relevant papers.

\section{Gas velocity distribution in collisions with ions}\label{sec:gasvelocitydistribution}

When ions collides with gas particles, special attention must be given to the statistical properties of gas particles which collide with a particular ion. Naively, one would think that statistical properties of gas particles which collide with an ion of a given velocity are independent of that velocity and of the cross section. This is not entirely correct, as we already explained in the introduction. However, it has been customary in literature \cite{VS95,YH94} to sample gas
velocities directly from the gas velocity distribution function for use in MC simulations. This leads to inaccuracy especially for small energies of particles.

The error in sampling velocities from the gas velocity distribution function $F(\mathbf{u})$ and from applying them in the collision kinematics may be
small or compensated by other effects. Still, this is a systematic error and one can never be sure how big the uncertainty is and whether it may become important under certain circumstances. We will see later that such
a procedure leads to results that are obviously wrong: at zero field and when the gas is described by the MVDF, the mean ion energy is smaller than the gas thermal energy and the ratio $eD_T/(kT\mu)$ is not equal one, as it should be (see Fig.~\ref{slikatrcoef}).

While most authors use the cold gas approximation and study mainly higher energies, some have attempted to obtain a more accurate method of calculating the distribution function of the collision frequency in their calculations \cite{LD04,RS03,NW99}.

The probability density per unit time for the collision of an ion of velocity $\mathbf v$ with a gas of velocity $\mathbf u$ is given by
\begin{align}
\label{gustinaverovatnocesudara} p(\mathbf v,\mathbf
u)=\frac{n\sigma\left(|\mathbf v-\mathbf u|\right)|\mathbf v-\mathbf
u|F(\mathbf u)}{\nu(v)},
\end{align}
where $F(\mathbf u)$ is the gas velocity distribution function and
$\nu(v)$ is the collision frequency, see
equation (\ref{nidef}). This is a conditional probability. The velocity of the gas particle $\mathbf{u}$ depends on the ion velocity $\mathbf{v}$ as well as on the total cross section.

A randomly chosen gas velocity should be chosen according to the probability density given by \ref{gustinaverovatnocesudara}. The method for the gas velocity sampling generally depends on the shape
of the cross section. A possible way to do it is to use the rejection method \cite{D86} which is easy for implementation, but is often not
efficient. Nevertheless, it is very hard to develop an efficient and quite general method for sampling for a wide range of possible cross section shapes, so the rejection method is very often the method of choice. For some special cases, like the constant cross section, one could furthermore transform equation (\ref{gustinaverovatnocesudara}) that we considered in the next subsection.

Two different implementations of the rejection method for
(\ref{gustinaverovatnocesudara}) are obvious. The first one is as follows. Let $q_{max}$ be the
maximum of $\sigma\left(|\mathbf v-\mathbf u|\right)|\mathbf
v-\mathbf u|$ for all allowed $\mathbf v$ and $\mathbf u$ (which are
specified in the beginning of simulation). This maximum can be
chosen as
\begin{align}
q_{max}=\sqrt{27w^2+6vw+v^2}\cdot\mathrm{max}_{\mathbf v,\mathbf
u}\sigma\left(|\mathbf v-\mathbf u|\right).
\end{align}
Sampled velocity $\mathbf u'$ from the MVDF is accepted as the
background velocity if
\begin{align}
\label{rejectionccs} \sigma\left(|\mathbf v-\mathbf
u'|\right)|\mathbf v-\mathbf u'|>rq_{max}
\end{align}
is fulfilled ($r$ is a uniform random number from $[0,1]$); otherwise
we sample another velocity from the MVDF until the condition
(\ref{rejectionccs}) becomes fulfilled. This method is good when the
cross section resembles a constant cross section and does not have sharp maxima. On the other hand, when the cross section has a sharp
maximum for a very narrow interval of velocities
(e.~g.~$\sigma(v)\sim 1/v$), the previous method is very inefficient and we may use another estimation for $q_{max}$ simply as
$\mathrm{max}_{\mathbf v_r}\sigma(\mathbf v_r)v_r$, where $v_r$ goes
over all allowed relative velocities (which are specified at the
beginning of simulation).

\subsection{Constant cross section}
\label{ccs}

In this subsection we will practically demonstrate artifacts of not using the expression (\ref{gustinaverovatnocesudara}) for determination of background gas velocities when they collide with an ion of a given velocity.

In the case of a constant cross section $\sigma_0$, expression (\ref{gustinaverovatnocesudara}) can be integrated analytically, and
the probability density per unit time for the collision of an ion of
velocity $v$ with a gas of velocity $u$ is:
\begin{align}
&p_{<}(v,u)=\frac{n\sigma_0}{\nu(v)}
{\frac{4}{3\sqrt\pi}\exp\left(-\frac{u^2}{w^2}\right)\frac{u}{w}\left(3\frac{u^2}{w^2}+\frac{v^2}{w^2}\right)},\\
&p_{>}(v,u)=\frac{n\sigma_0}{\nu(v)} \frac{4w}{3\sqrt\pi
v}\exp\left(-\frac{u^2}{w^2}\right)\frac{u^2}{w^2}\left(\frac{u^2}{w^2}+3\frac{v^2}{w^2}\right),
\end{align}
where $p_{<}(v,u)$ applies for velocities $v<u$ and $p_{>}(v,u)$ for $v>u$ and $\nu(v)$ for constant cross section is given by (\ref{nivhardfinal}).

\begin{figure}[t]
\centering
\includegraphics[width=\linewidth]{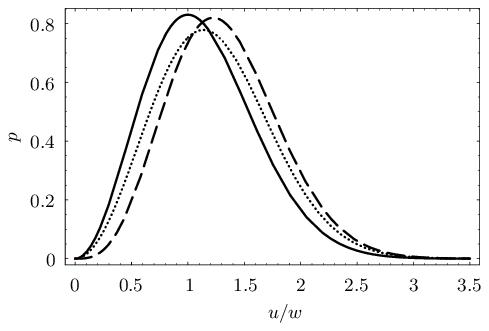}
\caption{Probability density of the sampled gas velocity $u$: the
case of zero ion velocity $v=0$ is shown by a dashed line, the case $v=w$ is shown by a
dotted line. MVDF is represented by a solid
line. When one samples gas velocities in an approximate way from MVDF, on average one always samples smaller velocities than using the correct
formula, which does not lead to the thermalization of ions at vanishing
electric field, as confirmed in the next section.}\label{slikamaksvelvstacan}
\end{figure}
\begin{figure}[t]
\centering
\includegraphics[width=\linewidth]{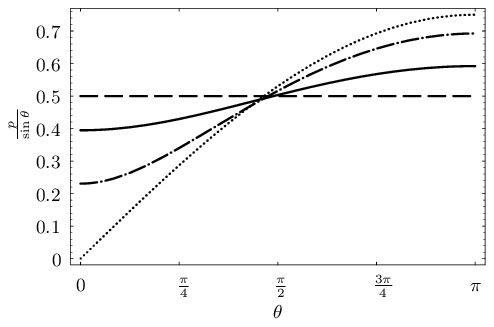}
\caption{Probability density of the azimuthal angle $\theta$ in the
case: $v=0$ -- dashed line, $v=w$ -- dotted line, $v=5w$ or $w=5v$
-- solid line, $v=2w$ or $w=2v$ -- dash-dotted
line.}\label{slikaugao}
\end{figure}

In Fig.~\ref{slikamaksvelvstacan} we show the probability density of
the sampled gas velocities for two different ion velocities in
comparison to the standard Maxwell's velocity distribution function.
We see that, with an increasing ion velocity, we approach the MVDF.
For ion velocities of the order of the  thermal gas velocity $w$ and
smaller, sampled gas velocities are considerably shifted towards
higher velocities (in respect to the MVDF).

Angular distribution may also be obtained from expression
(\ref{gustinaverovatnocesudara}). It is easy to see that the polar
angle distribution is uniform, but the azimuthal angle distribution
normalized by $\sin\theta$ is nonuniform. Azimuthalal angular
dependence for a given ion velocity $v$ and gas velocity $V$ is
\begin{align}
p(\theta)=\frac{3vV\left(v^2+V^2-2vV\cos\theta\right)^{1/2}}{(v+V)^3-|v-V|^3},
\end{align}
where $\cos\theta=\mathbf v \mathbf V/(vV)$. In
Fig.~\ref{slikaugao} we see that the  uniform (isotropic) normalized
azimuthal angle distributions correspond to the case of zero ion
velocity and also to the case when ion velocity is much higher than
the thermal velocity of the background gas. In the case $v=w$ we
have the highest anisotropy and angles closer to $\pi$ are more
probable. As the ratio $v/w$ increases or decreases the angular
distribution approaches the uniform case. Had we sampled the gas
velocity from the gas velocity distribution function, we would have
obtained uniform azimuthal distribution, since the gas velocity
distribution function is uniform in the space. We may conclude that all deviations made by not using the probability density (\ref{gustinaverovatnocesudara}) lead to systematic errors, both in magnitudes of sampled gas velocities and angular distributions.

\section{Mixtures of gases}

If the background gas of density $n$ is a mixture of $k$ gases with
relative densities $\alpha_i,\;\sum_{i=1}^k \alpha_i=1$, the
collision frequency (\ref{def_kolizionav}) becomes
\begin{align}
&\nu(v)=\sum_{i=1}^{k}\alpha_i \nu_i(v),\\
\label{nismesa} &\nu_i(v)=n\int\!\!\!\int\!\!\!\int\sigma_i
\left(|\mathbf{v}-\mathbf{u}|\right) {|\mathbf{v}-\mathbf{u}|}
F_i(\mathbf{u})\dif\mathbf{u},
\end{align}
where $F_i(\mathbf u)$ is the gas velocity distribution (which has
the same functional form for all gas components if the background
gas is in equilibrium; the difference is only in the value of the
thermal velocity due to different masses) and $\sigma_i$ is the
ion-neutral cross section of the component $i$. Expression
(\ref{nismesa}) can be easily calculated having in mind the
previously described procedure for a pure background gas.

Probability density per unit time for collisions of ions of velocity
$\mathbf v$ with a gas particle of velocity $\mathbf u$ is given by
\begin{align}
&p(\mathbf v,\mathbf u)=\sum_{i=1}^{k}w_i\frac{n\sigma_i
\left(|\mathbf{v}-\mathbf{u}|\right) {|\mathbf{v}-\mathbf{u}|}
F_i(\mathbf{u})}{\nu_i(v)},\\
&w_i=\frac{\alpha_i\nu_i(v)}{\nu(v)}.
\end{align}
Sampling a uniform number in accordance to weighting factors $w_i$,
we first determine the constituent of the mixture with which the ion
of velocity $v$ collides, and then we sample the velocity $u$ of
that constituent from the appropriate probability density
\cite{D86}. Using the equations from this subsection and the results of the part of the paper it is a straightforward procedure to consider thermal gas effects in mixtures of background gases.

\section{Some details of an MC simulation}

In this section we provide some details of an MC simulation that is intended to be used as a solver for the BE (\ref{BE}). While the procedure is well known and documented in literature\cite{N00}, some details are in order. First we derive the expressions for collision kinematics, since we have not found a compact and simple formulas in literature that are valid for particle collision of arbitrary masses and inelastic processes. Afterwards we briefly describe the procedure of simulation.

\subsection{Collision kinematics}

We consider two particles of masses $m$ and $M$ which collide. We assume that they have velocities $\mathbf v$ and $\mathbf V$, respectively with respect to the laboratory coordinate system with
$\mathbf e_x$, $\mathbf e_y$, and $\mathbf e_z$. Assuming the momentum
is conserved during collisions, the postcollision velocities are
\begin{align}
\label{v1prim} \mathbf v'&=\mathbf v-\frac{M}{m+M}(\mathbf
g-\mathbf g'),\\
\label{v2prim} \mathbf V'&=\mathbf V+\frac{m}{m+M}(\mathbf
g-\mathbf g'),
\end{align}
where
\begin{align}
&\mathbf g'=g'\mathbf e_{g'}=\mathbf v'-\mathbf
V'=\sqrt{(\mathbf v-\mathbf V)^2-\frac{2E}{\mu}}\mathbf e_{g'},\\
&\mu=\frac{mM}{m+M},\\
&\mathbf v_c=\frac{m\mathbf v+M\mathbf
V}{m+M}\\
&\mathbf g=\mathbf v-\mathbf V,
\end{align}
while $E$ is the inelastic
threshold. In the case of isotropic scattering $\mathbf e_{g'}$ is
the isotropic unit vector which in simulation may be chosen using
the algorithm proposed in \cite{K70}.

\begin{figure}
\centering
\includegraphics[width=0.9\columnwidth]{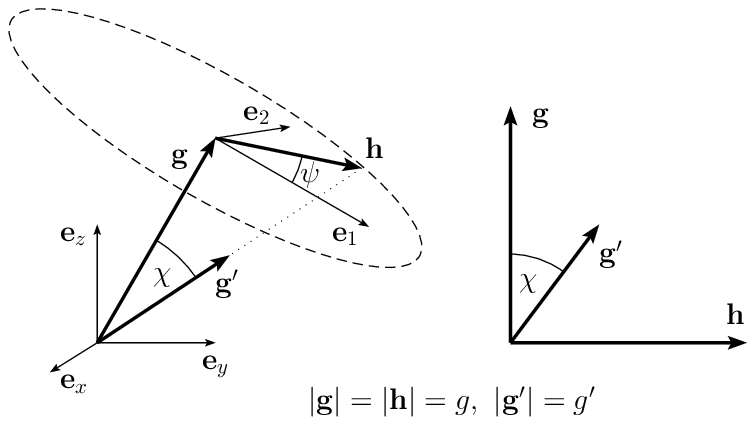}
\caption{Scattering of two particles: the relative velocity  before collision is $\mathbf{g}$ and after it is $\mathbf{g}'$. The angle between the two velocities $\chi$ is determined by the differential cross section, see the main text for more details.}\label{slikacoordinates}
\end{figure}

In the case of anisotropic scattering let $\chi\ (0\leq\chi\leq\pi)$
denote the azimuthal and $\psi\ (0\leq\psi\leq2 \pi)$ the polar
angle of the relative particle velocity after collision with respect
to the velocity before collision, see Fig.~\ref{slikacoordinates}. We have $\mathbf{g}\mathbf{g}'=g g'\cos\chi$. While the polar angle is completely undetermined and in the simulation is determined as a uniform random number from the interval $[0,2\pi)$, the azimuthal angle $\chi$ is determined by the differential cross section. In the simulation, it is obtained as a solution of the equation
\begin{align}
\label{jednacinanaizrasejanje} \frac{\int_0^\chi
\sigma^d(v,\chi)\sin\chi\dif \chi}{\int_0^\pi
\sigma^d(v,\chi)\sin\chi\dif \chi}=r,
\end{align}
where $r$ is a uniform random number from $[0,1]$, and
$\sigma^d(v,\chi)$ is the differential cross section.

Having the angles $\psi$ and $\chi$, it is not a difficult task to determine $\mathbf{g}'$ as a function of the angles and the relative velocity before the collision $\mathbf{g}$. Using that the unit vectors orthogonal to $\mathbf{e}_g=(g_x,g_y,g_z)/g$ are $\mathbf{e}_1=\left(g_x g_z,g_y g_z,-g_\rho^2\right)/(gg_\rho)$ and $\mathbf{e}_2=\left(-g_y,g_x,0\right)/g_\rho$, we have $\mathbf{h}=g\mathbf{e}_h=g\cos\psi \mathbf{e}_1+g\sin\psi\mathbf{e}_2$ and $\mathbf{g}'=g'\cos\chi\mathbf{e}_g+g'\sin\chi\mathbf{e}_h$, see Fig.~\ref{slikacoordinates}. Therefore, the postcollision velocities read
\begin{align} \label{v1primaniz}
\mathbf v'=\mathbf v-\frac{M}{m+M}\left(\mathbf
g-\frac{g'}{g}(\mathbf h \sin\chi+\mathbf g\cos\chi)\right),\\
\label{v2primaniz} \mathbf V'=\mathbf
V+\frac{m}{m+M}\left(\mathbf g-\frac{g'}{g}(\mathbf h
\sin\chi+\mathbf g\cos\chi)\right),
\end{align}
where $\mathbf h$ has components
\begin{align}
h_x=&(g_xg_z\sin\psi-gg_y\cos\psi)/g_\rho,\\
h_y=&(g_yg_z\sin\psi+gg_x\cos\psi)/g_\rho,\\
h_z=&-g_\rho\sin\psi,\\
g_\rho=&\sqrt{g_x^2+g_y^2}.
\end{align}
Equivalent expressions for the case of purely elastic scattering ($E=0$) can be found in \cite{N00}.

\subsection{Flowchart of simulation}

In order to solve the BE (\ref{BE}) and determine the distribution function of ions $f(\mathbf{x},\mathbf{v},t)$ one follows the evolution of an ion (or certain number of them). Once the ions relax one performs measurement of their velocities, positions, from which one produces wanted information (like the energy, the diffusion coefficients, the distribution function, etc.) that have to be averaged over many uncorrelated measurements.

In the following we consider a single ion. Knowing the collision frequency (\ref{nidef}) as a function of  the ion velocity $v$, one (stochastically) determines the time of its evolution before a collision with the neutral gas particle takes place. The evolution in between collisions is purely classical determined by the external force. The probability density that an ion that starts evolution at moment $t=0$ will suffer a collision with a background gas particle after the time $t$ is given by \cite{N00}
\begin{align}\label{vers}
p(t)=\nu(t)\exp\left[-\int_0^t\dif t' \nu(t')\right].
\end{align}
Once the probability density for a collision is known, the time for ion evolution is fully determined. The most direct way to determine $t$ is by solving the equation
\begin{align}
\int_0^{t}\dif t'p(t')=r,
\end{align}
where $r$ is a uniform random number from $[0,1]$. Since the ion velocity is a function of time, the integration of the previous equation is possible. In practice one often applies different methods for sampling $t$ from (\ref{vers}). One of the most important ways, so called the null-collision method \cite{skullerud64}, is the simplest rejection method.\cite{D86}
After the time of free evolution $t$ the ion collides with a gas particle. At that moment we do need the velocity $\mathbf{u}$ of the gas particle that collides with the ion having the velocity $\mathbf{v}(t)$ and that is determined by the probability density (\ref{gustinaverovatnocesudara}). The method for sampling $\mathbf{u}$ from (\ref{gustinaverovatnocesudara}) is described in section \ref{sec:gasvelocitydistribution}. Knowing the precollision velocities, we determine the postcollision velocities in a way described in the previous subsection. Since we do not follow the evolution of gas particles, we actually need only the ion velocity after the collision.

\begin{figure}
\centering
\includegraphics[width=1\columnwidth]{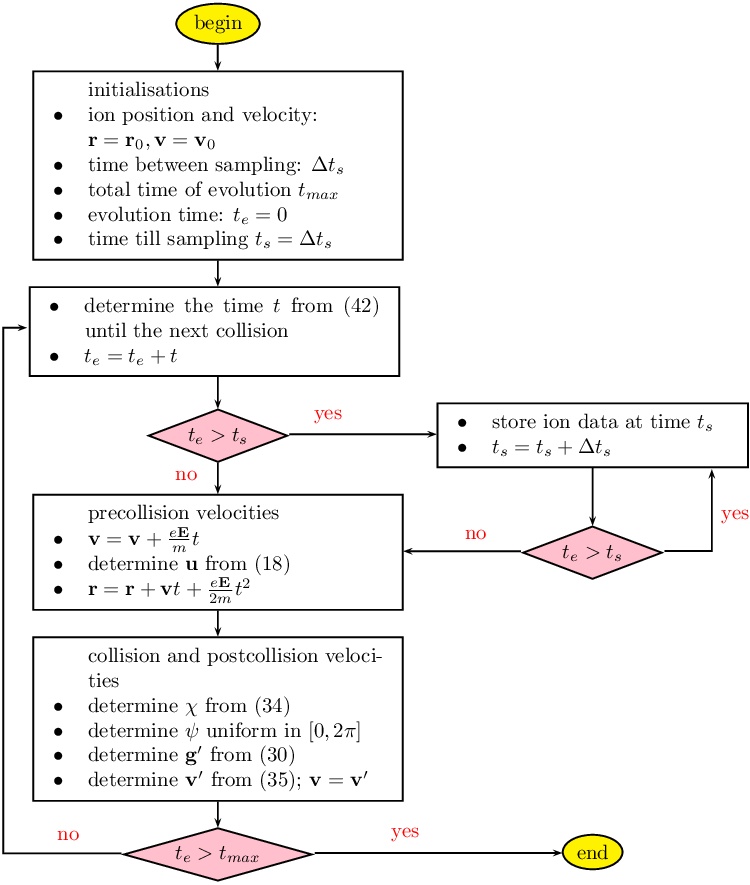}
\caption{Flowchart for the procedure of simulation.}\label{flowchart}
\end{figure}

The above procedure of evolution of a single ion and its collision with a gas particle is repeated over many collisions until the wanted precision of results is obtained. The flowchart is given in Fig.~\ref{flowchart}.

\section{Results of MC simulation}

We have developed a new MC simulation code for ion
transport in a background gas in an external electric field using the methods described in the previous sections.
The code was required to pass all the benchmark test calculations that are available for a zero gas temperature. We implemented in the code a procedure for collision frequency calculation and for sampling the gas velocity as described in this paper. The number of benchmark results that could be used to test the nonzero gas temperature calculations that are available in literature is very small. Thus we have tried to produce new values while at the same time repeating the available benchmark results.

There are several sources for benchmark calculations for ions
\cite{S73,PS98}, but we found that only reference \cite{WRN01} does not use the cold gas approximation while providing a range of results that may be used to test the code.

Numerical results of simulations  in the case of a constant cross section, for two ion-gas mass ratios are presented in Table~\ref{rezultatm1m4}. There we compare our results to the BE results of White et al.~\cite{WRN01}. In the case of a
small ion-gas mass ratio and a constant cross section the MC simulation takes a long time because of very slow relaxation
of ion transport properties (very inefficient energy transfer from
the ion to the gas particle in elastic scattering because of the low
mass ratio), and our results are less accurate than the results for
equal masses. Nevertheless, our results are in excellent agreement
with the results from the numerical solution of the BE \cite{WRN01}.

\begin{table*}
\caption{Results for the ions with a hard sphere cross section model
for $m_{gas}=4\,\mathrm{amu}$, $T=293\,\mathrm{K}$,
$E/n=1\,\mathrm{Td}$, $\sigma_0=6\cdot
10^{-20}\,\mathrm{m}^2$ The first column are the results for $T=0$ for comparison.\label{rezultatm1m4}}
\begin{ruledtabular}
\begin{tabular}{lccc|cc}
&\multicolumn{3}{c}{$m_{ion}=m_{gas}$}&\multicolumn{2}{c}{$m_{ion}=10^{-4}m_{gas}$}\\
&$T=0$&$T=293\ \mathrm{K}$&Results from [43]&$T=293\ \mathrm{K}$&Results from [43]\\ \hline
$\varepsilon[\mathrm{eV}]$ &$0.01962$& $0.04271$ & $0.04271$& $0.7332$ & $0.73324$\\
$W~[10^2 \mathrm{ms}^{-1}]$ &$7.270$& $3.367$ &$3.368$ & $56.27$ &$56.187$ \\
$nD_L~[10^{22}(\mathrm{ms})^{-1}]$ &$0.233$& $0.885$ & $0.884$\footnotemark[1] & $155$ & $158.27$ \\
$nD_T~[10^{22}(\mathrm{ms})^{-1}]$ &$0.338$& $0.894$ & $0.894$\footnotemark[1] & $315$ & $313.20$  \\
$T_T~[10^2\mathrm{K}]$ &$0.5664$& $3.074$ & $3.074$ & $56.70$ & $56.717$ \\
$T_L~[10^2\mathrm{K}]$ &$0.8762$& $3.220$ &$3.220$& $56.72$ &$56.727$\\
\end{tabular}
\end{ruledtabular}
\footnotetext[1]{Note an error for a factor $10$ in the original data
from the paper \cite{WRN01}.}
\end{table*}

\begin{table*}
\caption{Benchmark results for Ar$^+$ ions in Ar for
$m=2$.\label{rezultatar+ar}}
\begin{ruledtabular}
\begin{tabular}{lllllll}
&\multicolumn{3}{c}{$T=77\,\mathrm{K}\qquad
w=179.0\,\mathrm{ms^{-1}}$}&\multicolumn{3}{c}{$T=293\,\mathrm{K}$\qquad
$w=349.2\,\mathrm{ms^{-1}}$}\\ $E/n\,[\mathrm{Td}]$ &
$\varepsilon\,[\mathrm{meV}]$ & $v\,[\mathrm{ms^{-1}}]$ &
$nD_{L}[10^{22}(\mathrm{ms})^{-1}]$ & $\varepsilon\,[\mathrm{meV}]$
& $v\,[\mathrm{ms^{-1}}]$ &
$nD_{L}[10^{22}(\mathrm{ms})^{-1}]$\\
\hline
$1  $ & $ 9.97$ & $ 5.98$ & $0.00400$ & $37.89$ & $  4.26$ & $0.0108$ \\
$5  $ & $10.32$ & $29.86$ & $0.00410$ & $38.07$ & $ 21.22$ & $0.0108$ \\
$10 $ & $11.39$ & $58.89$ & $0.00436$ & $38.64$ & $ 42.31$ & $0.0108$ \\
$16 $ & $13.41$ & $92.07$ & $0.00482$ & $39.77$ & $ 67.21$ & $0.0110$ \\
$18 $ & $14.23$ & $102.6$ & $0.00499$ & $40.26$ & $ 75.41$ & $0.0111$ \\
$20 $ & $15.12$ & $113.0$ & $0.00516$ & $40.80$ & $ 83.48$ & $0.0111$ \\
$30 $ & $20.25$ & $161.3$ & $0.00601$ & $44.13$ & $122.9$ & $0.0116$ \\
$35 $ & $23.16$ & $183.5$ & $0.00644$ & $46.16$ & $141.9$ & $0.0119$ \\
$40 $ & $26.24$ & $204.7$ & $0.00683$ & $48.38$ & $160.3$ & $0.0121$ \\
$45 $ & $29.46$ & $224.8$ & $0.00724$ & $50.78$ & $178.3$ & $0.0124$ \\
$50 $ & $32.79$ & $244.1$ & $0.00760$ & $53.35$ & $195.8$ & $0.0126$ \\
$60 $ & $39.74$ & $280.3$ & $0.00836$ & $58.89$ & $229.4$ & $0.0131$ \\
$65 $ & $43.33$ & $297.4$ & $0.00874$ & $61.83$ & $245.6$ & $0.0135$ \\
$70 $ & $46.99$ & $314.1$ & $0.00908$ & $64.88$ & $261.4$ & $0.0137$ \\
$80 $ & $54.45$ & $345.7$ & $0.00977$ & $71.25$ & $291.8$ & $0.0141$ \\
$90 $ & $62.12$ & $375.5$ & $0.0105 $ & $77.93$ & $321.0$ & $0.0147$ \\
$100$ & $69.93$ & $403.9$ & $0.0111 $ & $84.84$ & $348.9$ & $0.0153$ \\
$120$ & $86.02$ & $456.9$ & $0.0123 $ & $99.38$ & $401.7$ & $0.0163$ \\
$140$ & $102.5$ & $505.8$ & $0.0136 $ & $114.6$ & $450.9$ & $0.0174$ \\
$150$ & $110.9$ & $529.1$ & $0.0141 $ & $122.5$ & $474.4$ & $0.0178$ \\
$160$ & $119.3$ & $551.5$ & $0.0146 $ & $130.4$ & $497.2$ & $0.0184$ \\
$200$ & $153.9$ & $635.5$ & $0.0168 $ & $163.4$ & $582.8$ & $0.0202$ \\
$300$ & $244.4$ & $816.0$ & $0.0216 $ & $250.9$ & $767.2$ & $0.0246$ \\
$350$ & $291.1$ & $895.2$ & $0.0239 $ & $296.7$ & $848.4$ & $0.0267$ \\
$400$ & $338.8$ & $969.5$ & $0.0260 $ & $343.6$ & $924.2$ &$0.0288$
\end{tabular}
\end{ruledtabular}
\end{table*}

\begin{figure*}
\centering
\includegraphics[width=1\linewidth]{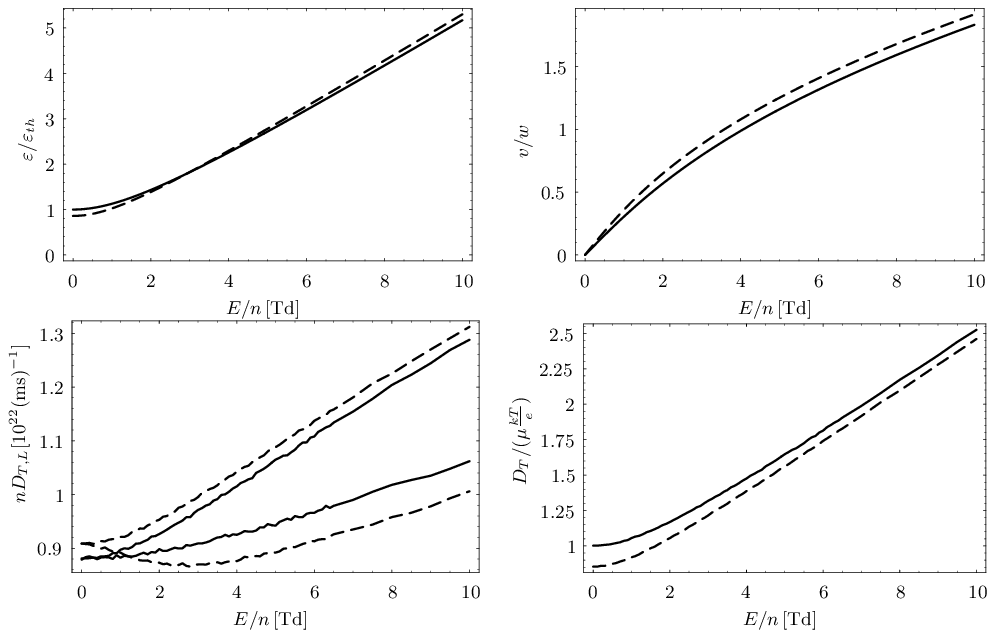}
\caption{Mean ion energy, drift velocity, longitudinal and
transverse diffusion coefficient and transverse diffusion
coefficient over mobility as a function of reduced electric field
for a hard sphere model for $m_{ion}=m_{gas}=4\,\mathrm{amu}$,
$T=293\,\mathrm{K}$, $\sigma_0=6\cdot 10^{-20}\,\mathrm{m}^2$ for
different sampling procedures of velocity of background gas: from expression
(\ref{gustinaverovatnocesudara}) -- solid line; from MVDF -- dashed
line. In the third figure, the upper curve corresponds to $nD_T$ and the lower to $nD_L$, i.e.~it is satisfied $D_T>D_L$.}\label{slikatrcoef}
\end{figure*}

In Fig.~\ref{slikatrcoef} we show results from the simulation for
different electric fields. We applied different samplings procedures
for the background gas: from the correct formula
(\ref{gustinaverovatnocesudara}) and directly from the gas velocity
distribution function (which is assumed to be MVDF). One can see
that thermal equilibrium (at zero field) cannot be achieved when the
background velocities are not sampled in a proper way: mean ion
energy is lower than the thermal energy $3kT/2$ and the
characteristic energy $eD_T/\mu$ is lower than $kT$.

\begin{figure}
\centering
\includegraphics[width=1\linewidth]{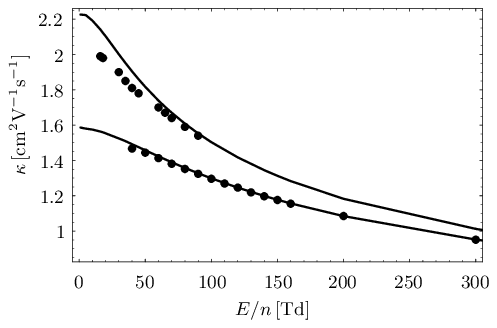}
\caption{Calculated reduced mobilities of Ar$^+$ ions in Ar at
$77\,\mathrm{K}$ (upper curve) and $293\,\mathrm{K}$ (lower curve).
The dots are experimental results from
\cite{H77,H78}. One can note that the mobilities decrease increasing the temperature.}\label{slikakapa}
\end{figure}

In Fig.~\ref{slikakapa} we show results for reduced mobilities for
Ar$^+$ ions in their parent gas using a set of cross sections from
\cite{P94}. One can see that measured values at $293\,\mathrm{K}$
\cite{H77} are in excelent agreement with our calculated values,
while measured values at $77\,\mathrm{K}$ \cite{H78} are slightly
lower than calculated. This may indicate the need to make some small
adjustments to the cross section in the low energy limit. We should
mention that measurements of Basurto et al.~\cite{BUAC00} agree very
well with those of Helm and with our calculations.

In Table \ref{rezultatar+ar} we tabulate the results for Ar$^+$ ions
in their parent gas which may be used as benchmark results for
verification of MC codes or other techniques which take
into account thermal effects of the background gas. We should
mention that relative differences between results for two choices of parameters $m=2$ and $m=2.3$ (defined in\cite{P94}) are less than one percent,
and only the results for $m=2$ are presented here.

%For $T=293\,\mathrm{K}$ the transverse diffusion coefficient is constant and has a value $nD_{T}=0.010\, %[10^{22}(\mathrm{ms})^{-1}]$ which is maybe strange (??) and should be addressed to the set of cross sections. However %at $T=77\,\mathrm{K}$ the transverse diffusion coefficient grows from $nD_{T}=0.004\,
%[10^{22}(\mathrm{ms})^{-1}]$ at $E/n=1\,[\mathrm{Td}]$ up to the
%value of $nD_{T}=0.006\, [10^{22}(\mathrm{ms})^{-1}]$ at
%$E/n=400\,[\mathrm{Td}]$.

In \cite{RS03} an approximate calculation of the collision frequency was carried out in an attempt to include properly  the thermal collisions. The difference of the collision frequency is appreciable as can be seen in Fig.~1 although it is an improvement on the cold gas approximation. If we compare our results to the data from the analytic formula for the drift velocity given in \cite{RS03}, the differences are up to 10$\%$ in the low $E/n$ range. Results for the hard sphere for T=0 K are included in Table I. These values are only for 1 Td and the mean energy would differ even more at lower $E/n$. The mobility already differs by a factor of two. As for argon, the effect of T=0 approximation can be estimated by comparing 77 K and 293 K results. The mean energy will continue to decrease going towards zero and the difference in mobility under those conditions in case of argon it could be as large as 30$\%$.

\section{Conclusions}

We have developed an accurate and efficient algorithm for
calculating the collision frequency in the case when thermal motion
of background gas cannot be neglected for a Maxwellian velocity
distribution of the background gas particles. This is required for
low energy simulations of ion transport using an MC
technique or for implementation of hybrid and particle in cell
codes \cite{B91,BKV98}. Exact formulas are derived for very general cross section dependencies on relative velocity. Also, an analytic
form of the correction terms in the collision frequency was found,
in the case of a large ratio of ion--gas thermal velocities. It was
emphasized that sampling of the gas velocity for a given ion
velocity (in MC simulations) should be done by using an
appropriate probability of collisions. Explicit analytic formulas
are obtained in the case of a constant cross section, and it was
shown that the velocity distribution function of the gas
particles which collide with an ion of a given velocity is
significantly different from a Maxwell distribution. The results
of our MC simulation for ion transport are in excellent
agreement with the numerical solution of the BE and
may be used as a benchmark for ion transport properties.

In a recent attempt to provide the data for some gases where data
were missing and to go beyond the basic Langevin theory, Nanbu and
coworkers have defined theory in terms of collision probabilities
\cite{NK95,DN98,TN05} rather than cross sections. It may be worth
the effort to include the presently proposed technique together with
their approach, especially since that theory is normalized by using
the mobilities which are usually available only in the thermal limit
or close to it.

It is worth noting that so far very accurate numerical techniques
for solutions of the BE have been used to study ion
transport \cite{KMV90,VK96,MM88}. Application of MC
simulations  have not been as widespread for ions as they have been
for electrons. This to some degree is due to the problems in
representing thermal collisions which are important over a much
broader range of $E/n$. The present technique has already been
applied in simulations of the transport of thermal ions
\cite{PJSRR08,PRJSMMU07}. At the same time Boltzmann techniques
mostly use interaction potentials to communicate the results to the
scientific community. While it is possible to convert interaction potentials to cross
sections it is difficult and may not lead to unique results \cite{viehlandcpc+10}.  Plasma modelers may also have a difficulty to appreciate the applicability of certain sets of data for potentials over a
wide range of mean energies and values of $E/n$.  Thus it seems
better to apply MC simulations to convert the transport
data to cross sections as such codes may be directly compatible with
plasma models or even more so directly applicable to build hybrid
models that treat ion transport by MC simulations
\cite{DHK07,DHK06,D01,GBG03,GBG04,MHMDP03}.

In general all MC techniques for modeling swarms \cite{PS+07,PD+09} and discharges  would benefit from implementing our procedure. The same is true for the hybrid codes \cite{Kushner09}, particle in cell-MC models \cite{Donko11} and calculation of the data for global models \cite{LiuBIRK10}.  Special cases where it is important to include proper treatment of collision frequency in the limit of thermal energies are for example modeling of the afterglow or flowing afterglow \cite{Kutasi07,LoureiroGSPL11,PintassiglioGGR10,KutasiSPLM}   for various applications, border between the sheath and the bulk of glow discharges (both in direct- and in radiofrequent-current regimes), microwave discharges,  high and atmospheric pressure discharges \cite{LoH11,sakiyamagraves} including plasmas for medical applications, coronas \cite{DutenRAVH} and also for gaseous elementary particle detectors \cite{FonteP10}.  While surprising behavior of plasma bullets \cite{NiemiWSGOC11} and motion of ionization fronts implies high field regions these have to be modeled within the field free background and for relaxation of the distribution function one needs to implement a technique similar to the one presented in this paper. Thermalization of charged particles in gases, dielectrics, living tissue and atmosphere is of special importance for emerging applications.

The technique described here has been applied in a number of situations during the preparation of this paper. Several publications dealing with different problems were published. While most of them cover the thermal region relatively little \cite{PJSRR08,PRJSMMU07,SU+10}, the results indicate that the code has passed all other tests including those for reactive collisions, for anisotropic collisions and tests against hard sphere benchmark for very different mass ratios. The code has also been used to show that for ions there are kinetic phenomena dictated by non-conservative collisions \cite{PRJSMMU07,PS+07,PD+09}.

An additional test came from recent studies of thermalization of positrons and positronium \cite{Psthermal} where  direct sampling of the relative velocity from the gas distribution function proved to give incorrect limits, most importantly the mean energy. Only when the procedure described in the present paper is employed, the correct limits have been achieved \cite{Psthermal,WT+11}. Thus the present results will prove useful in modeling of the effects of positrons in positron emission tomography and positron therapy \cite{Garciaetal2011}.

The proposed procedure, while somewhat more complex than the basic MC procedure of \cite{LB77} is not significantly more demanding in terms of computation time. The only difference is the sampling of the background velocities using the distribution (\ref{gustinaverovatnocesudara}). In general the needs for data for modeling of plasmas including the low energy limit \cite{NapartovichK2011,MakabeT2011,PS+07,PD+09}  show that this procedure should be implemented in all the MC codes when accurate data are required including the ability to represent spatial and temporal dependencies of the field.
It is important to note that procedure in \cite{LB77} or cold gas approximation become sufficiently accurate at energies that are several times larger than the thermal energy, in the case of ions in their parent gas. The reason why the problem with the sampling was not found some years ago was due to the fact that only a few studies were made that focused on very low energies and also because statistics of simulations was too poor to notice the effects.

\begin{acknowledgments}
Work on this paper was supported by OI 171037 and II41011 (ZLP) and in part by ANR grant 09-BLAN-0097-01/2 (ZR). The authors are grateful to S.~Vrhovac and M.~\v{S}uvakov for discussions and useful suggestions.

\end{acknowledgments}

\appendix
\section{}
Consider an integral of the form
\begin{align}
\label{app1} I=\frac{1}{\sqrt\pi w}\int_{-\infty}^{+\infty}f(x)
\exp\left[-\frac{(x-v)^2}{w^2}\right]\dif x.
\end{align}
Let $f(x)$ be a function which is continuous and repeatedly
differentiable at $x=v$, and furthermore the integral in
(\ref{app1}) converges. We transform this integral by the
substitution $t=(x-v)/w$:
\begin{align}
\label{app2} I=\frac{1}{\sqrt\pi}\int_{-\infty}^{+\infty}f(v+wt)
\exp\left(-t^2\right)\dif t.
\end{align}
After expanding the function $f(v+wt)$ into Taylor series
\begin{align}
f(v+wt)=\sum_{n=0}^{\infty}\frac{(wt)^n}{n!}f^{(n)}(v),
\end{align}
using the formula
\begin{align}
\label{app3} \int_{-\infty}^{+\infty}\dif t t^{2n}
\exp\left(-t^2\right)=\Gamma\left(n+\frac{1}{2}\right),
\end{align}
we obtain the final formula
\begin{align}
\label{app4} I=\sum_{n=0}^{\infty}\frac{(w)^{2n}}{4^n
n!}f^{(2n)}(v).
\end{align}
In the limit of vanishing $w$, the integral (\ref{app1}) can be
calculated directly using representation of delta function
\begin{align}
\delta(v)=\frac{1}{\sqrt{\pi}}\lim_{w\rightarrow
0}\frac{\exp\left(-\frac{v^2}{w^2}\right)}{w}.
\end{align}
When one wants to calculate an integral of the form
\begin{align}
\label{app5} I'=\frac{1}{\sqrt\pi w}\int_{0}^{+\infty}f(x)
\exp\left[-\frac{(x-v)^2}{w^2}\right]\dif x
\end{align}
the situation is different. It is hard to obtain a general formula
for integral $I'$, but it is possible to calculate it in cases when
the condition $v\gg w$ holds. In that case the main contribution to
$I'$ comes from points $x$ which are in vicinity of $v$, so the
lower limit in integral (\ref{app5}) can be replaced by minus
infinity, and $I'$ is equal to the integral $I$ up to the
exponentially vanishing term:
\begin{align}
\label{app6}
I'=I+\mathcal{O}\left[\exp\left(-\frac{v^2}{w^2}\right)\right].
\end{align}

\bibliography{bibliography}
\end{document}